**Article title**

# Comprehensive and Spatially Detailed Passenger Vehicle and Truck Traffic Volume Data for the United States Estimated by Machine Learning

**Authors**

Brittany Antonczak[1], Meg Fay[1], Aviral Chawla[1,2], Gregory Rowangould[1]*

**Affiliations**

1. Transportation Research Center, University of Vermont, Burlington, VT 05405-0156, United States

2. Vermont Complex Systems Institute, University of Vermont, Burlington, VT 05405-0156, United States

**Corresponding author's email address**

Gregory.rowangould@uvm.edu

**Keywords**

Traffic Volume, Truck Traffic, Heavy-Duty Vehicles, Medium-Duty Vehicles, HPMS, Average Annual Daily Traffic, Traffic Density

**Abstract**

The Highway Performance Monitoring System, managed by the Federal Highway Administration, provides data on average annual daily traffic volume across roadways in the United States, but it has limited representation of medium- and heavy-duty vehicle traffic on lower-volume roadways that are not part of the national highway system. This gap limits research and policy analysis on the community impacts of truck traffic, especially concerning air quality and public health. To address this, we use random forest regression to estimate medium- and heavy-duty vehicle traffic volumes on network links where these data are missing. The result is a comprehensive vehicle traffic dataset that covers 85.2% of public roadways in the United States. From these data, we also calculate traffic density values for each census block and vehicle class that can serve as a high-resolution surrogate for traffic-related air pollution exposure in public health studies and policy analysis. Our high-resolution spatial data products are rigorously validated and provide a more complete representation of truck traffic than any existing publicly available dataset. These datasets are valuable for transportation planning, public health research, and policy decisions aimed at understanding and mitigating the effects of truck traffic on communities that are disproportionately exposed to air pollution from vehicle traffic.

# Estimated Roadway Segment Traffic Data by Vehicle Class for the United States: A Machine Learning Approach

Brittany Antonczak[1], Meg Fay[1], Aviral Chawla[1,2], Gregory Rowangould[1]

1. Transportation Research Center, University of Vermont, Burlington, VT 05405-0156, United States

2. Vermont Complex Systems Institute, University of Vermont, Burlington, VT 05405-0156, United States

Corresponding Author: Gregory Rowangould (gregory.rowangould@uvm.edu)

**Abstract**

The Highway Performance Monitoring System, managed by the Federal Highway Administration, provides essential data on average annual daily traffic across U.S. roadways, but it has limited representation of medium- and heavy-duty vehicles on non-interstate roads. This gap limits research and policy analysis on the impacts of truck traffic, especially concerning air quality and public health. To address this, we use random forest regression to estimate medium- and heavy-duty vehicle traffic volumes in areas with sparse data. This results in a more comprehensive dataset, which enables the estimation of traffic density at the census block level as a proxy for traffic-related air pollution exposure. Our high-resolution spatial data products, rigorously validated, provide a more accurate representation of truck traffic and its environmental and health impacts. These datasets are valuable for transportation planning, public health research, and policy decisions aimed at mitigating the effects of truck traffic on vulnerable communities exposed to air pollution.

**Background & Summary**

The Highway Performance Monitoring System (HPMS), managed by the Federal Highway Administration (FHWA), provides critical average annual daily traffic (AADT) data across most public roadways in the United States[1]. The HPMS dataset is a comprehensive, publicly available resource based on annual state-collected traffic counts across highways, arterials, and collectors. Using a stratified random sampling method, the data typically exhibits less than 10% error for at least 80% of roadway links[2]. While the dataset offers valuable insights into national transportation patterns, it suffers from sparse representation of medium-duty vehicle (MDVs) and heavy-duty vehicle (HDVs) traffic volume on non-interstate roadways. The availability of MDV and HDV data varies across states, reflecting differences in traffic counting equipment and reporting standards. HPMS data is vital for federal and state transportation planning, policy-making, and performance evaluation[1]. It informs key reports to Congress, including the biennial Condition and Performance Report, and supports apportionment of surface transportation funds under the Fixing America's Surface Transportation Act (FAST Act) [3–5]. The dataset also guides decision-making by Metropolitan Planning Organizations (MPOs) on air quality conformity and infrastructure investments[6]. Furthermore, HPMS data is used in government publications such as the FHWA Highway Statistics Series, as well as in assessing roadway safety and conditions[7].

The HPMS AADT data serves a broad spectrum of users beyond its primary applications by transportation planning agencies. The U.S. Environmental Protection Agency (EPA) use the HPMS data in MOVES (MOtor Vehicle Emission Simulator) to estimate emissions and assess air quality impacts, feeding into the National Emissions Inventory (NEI) for comprehensive air quality studies[8,9]. Social equity and environmental justice screening tools, such as the EPA's EJScreen, leverage HPMS AADT data to analyze transportation impacts on vulnerable communities[10]. In academia, the HPMS dataset is a vital publicly available resource for researching a variety of transportation-related topics. It provides essential data for assessing infrastructure conditions, analyzing traffic speed and congestion, and studying road safety and factors contributing to crashes[11–16]. Additionally, the data is used by researchers to evaluate sustainability and resiliency of U.S. transportation systems in the face of climate change, air pollution exposure, and extreme weather events[17–22]. It has been used in multiple academic studies on national exposure characterization and equity[18,19,21]. By analyzing traffic density and proximity to roadways—both correlated with elevated emissions of pollutants such as particulate matter (PM) and nitrogen oxides (NOx)—researchers have identified disparities in exposure among racial and socioeconomic groups[18,19,23,24]. Furthermore, the data is pivotal in calibrating travel demand models, optimizing transportation networks, and validating passive mobility data sources like Streetlight[22,25–27]. These diverse applications highlight the essential role of the HPMS in shaping policy, planning, and research across various sectors, extending well beyond traditional transportation engineering.

While the HPMS breaks out single-unit truck and bus (i.e., MDV) AADT and combination truck (i.e., HDV) AADT for all segments of the national highway and interstate system, states are only required to provide a random sample of MDV and HDV AADT for other roadway types. This sampling can limit the application of the HPMS in the wide range of interests in research and planning around truck traffic. The transportation of goods via trucks is vital to maintaining and growing strong local, regional, and national economies. It also poses significant transportation planning challenges[28]. Medium- and heavy-duty vehicles are a source of congestion, traffic safety concerns, and infrastructure damage[28–30]. Additionally, these trucks are often diesel-powered, raising air quality and public health concerns, contributing to greenhouse gas (GHG) emissions, and are implicated in environmental justice and equity studies as a large contributor to the disproportionate air pollution burden that lower income and communities of color experience[30–33]. Studies investigating these critical impacts on communities require detailed and accurate data to advance community-level research.

Few studies have addressed the data gap surrounding medium- and heavy-duty vehicles. FHWA research estimating lane volumes for pavement performance has calculated truck AADT by multiplying the total AADT by the percentage of each vehicle type[16]. As outlined in the state-level HPMS Vehicle Summaries Catalog, this percentage is derived as the average percentage of all sampled roads sharing the same roadway functional system, vehicle type, and urban-rural designation[7]. The U.S. EPA has developed a method for estimating the vehicle miles traveled (VMT) by road type and vehicle class, which serves as the default on-road vehicle activity in MOVES. The EPA estimates county-level VMT by road type and vehicle class by multiplying county-level VMT by the percentage of VMT for each vehicle class and road type, as reported in the Highway Statistics Series[8,34]. While these estimates serve their intended purpose, they do not provide roadway segment-level data that would be valuable for a wide range of policy, planning, and research applications.

Research and policy analysis related to truck traffic—critical for infrastructure, air quality, and public health— are hindered by incomplete roadway segment-level data on MDV and HDV traffic. Antonczak et al. (2023) previously estimated MDV and HDV traffic volumes using linear regression, revealing the disproportionate exposure to these vehicle types in communities of color and lower-income populations[18]. To further address this research gap, we improve upon linear regression by using random forest regression, which better handles complex interactions and nonlinear relationships, resulting in more accurate and comprehensive estimates. By leveraging available traffic data and road characteristics from the HPMS, our model estimates MDV and HDV AADT for areas where such data are unavailable, creating a more comprehensive national dataset. We then calculate traffic density at the census block level which serves as a proxy for exposure to traffic related air pollution (TRAP). Our high-resolution spatial data products—including road link-level traffic volumes and census block traffic density estimates—are rigorously validated, ensuring their reliability and robustness. These datasets provide a crucial foundation for future transportation planning, public health research, and policy decisions aimed at mitigating the adverse impacts of truck traffic on communities.

## Methods

### Data Overview and HPMS Dataset Description

We used vehicle activity data from the 2018 HPMS, a comprehensive dataset managed by the U.S. FHWA[1]. The dataset was obtained directly from the Bureau of Transportation Statistics (BTS) in December 2023 and represents the most detailed spatial data on vehicle traffic available for our study area, covering 50 U.S. states and Washington, D.C., as of the 2018 calendar year. The postprocessed dataset includes approximately 6.5 million rows, each representing a unique road link, and covers approximately 400,000 lane-kilometers of the U.S. highway system. The dataset records 12.4 billion vehicle-kilometers traveled (VKT) across six functional classifications of roadways. **Table 1** provides a breakdown of lane-kilometers and VKT by functional classification and vehicle type, offering a detailed view of traffic dynamics across the U.S. highway system. These estimates, based on state-collected traffic counts, form the foundation of our analysis, which focuses on estimating MDV and HDV traffic volume on links in the HPMS where these data are unavailable. In the HPMS dataset, around 28% of road links (or 47% of lane-kilometers) lack MDV and HDV AADT data. This data gap is particularly notable on lower-volume roads and in rural areas where traffic counting may be less frequent or standardized across states.

**Table 1.** Breakdown of Lane-Kilometers and Vehicle-Kilometers Traveled (VKT) by Functional Classification and Vehicle Type

| Functional Classification | | Lane-Kilometers (in thousands) | VKT (in millions) | | |
|---|---|---|---|---|---|
| **Code** | **Description** | | **Total** | **MDV** | **HDV** |
| **1** | Interstate | 418.6 | 4,052 | 155.0 | 452.3 |
| **2** | Other Freeways & Expressways | 147.6 | 1,380 | 48.89 | 57.17 |
| **3** | Other Principal Arterial | 751.7 | 2,954 | 122.9 | 134.5 |
| **4** | Minor Arterial | 891.1 | 2,357 | 44.03 | 31.57 |
| **5** | Major Collector | 1711 | 1,568 | 24.24 | 14.41 |
| **6** | Minor Collector | 60.09 | 76.98 | 1.429 | 0.515 |
| **Total** | | 3,980 | 12,389 | 396.5 | 690.6 |

*Note. VKT = vehicle-kilometers traveled; MDV = medium-duty vehicle; HDV = heavy-duty vehicle. Separate tabulations for medium-duty and heavy-duty vehicle vehicle-kilometers traveled are based on 2018 HPMS sampled annual average daily traffic values. The values represent the 50 U.S. states and Washington, D.C.*

**Data Preprocessing and Geospatial Analysis**

All spatial analysis was conducted within ArcGIS Pro version 4.0.2. First, the Repair Geometry tool was applied to detect and fix geometric issues, such as null or invalid geometries. Next, the data was subsetted to include only the 50 U.S. states and Washington, D.C. Road links were then attributed with county-level information by intersecting the HPMS road network with 2020 U.S. Census county boundaries[35].

The 2010 Urban Area Census Code (UACE) was initially used to classify road segments into urban, rural, and small urban areas in the HPMS dataset. However, inaccuracies in the UACE codes (e.g., missing or erroneous entries) were identified. To address these issues, the HPMS dataset was realigned with 2010 U.S. Urban Areas (UAC) Boundaries[36], correcting discrepancies in the classification. This adjustment modified approximately 14% of road links (representing 16% of lane-kilometers) and resulted in a new urban-rural classification field with values: '0' for rural, '1' for urban, and '2' for small urban. Distinguishing between urban and rural areas is crucial for accurate traffic volume estimation, as traffic patterns, road conditions, and vehicle behavior can differ significantly across these settings. This adjustment ensures more accurate traffic volume estimations for MDVs and HDVs in areas where data are unavailable while correcting discrepancies in the original UACE data.

To improve the quality of the analysis, road links lacking total AADT estimates or with AADT values lower than the sum of MDV and HDV AADT were excluded. This process resulted in the removal of approximately 4.8% of road links (around 11% of lane-kilometers). For road links with missing information on through lanes, a default of two lanes was assumed.

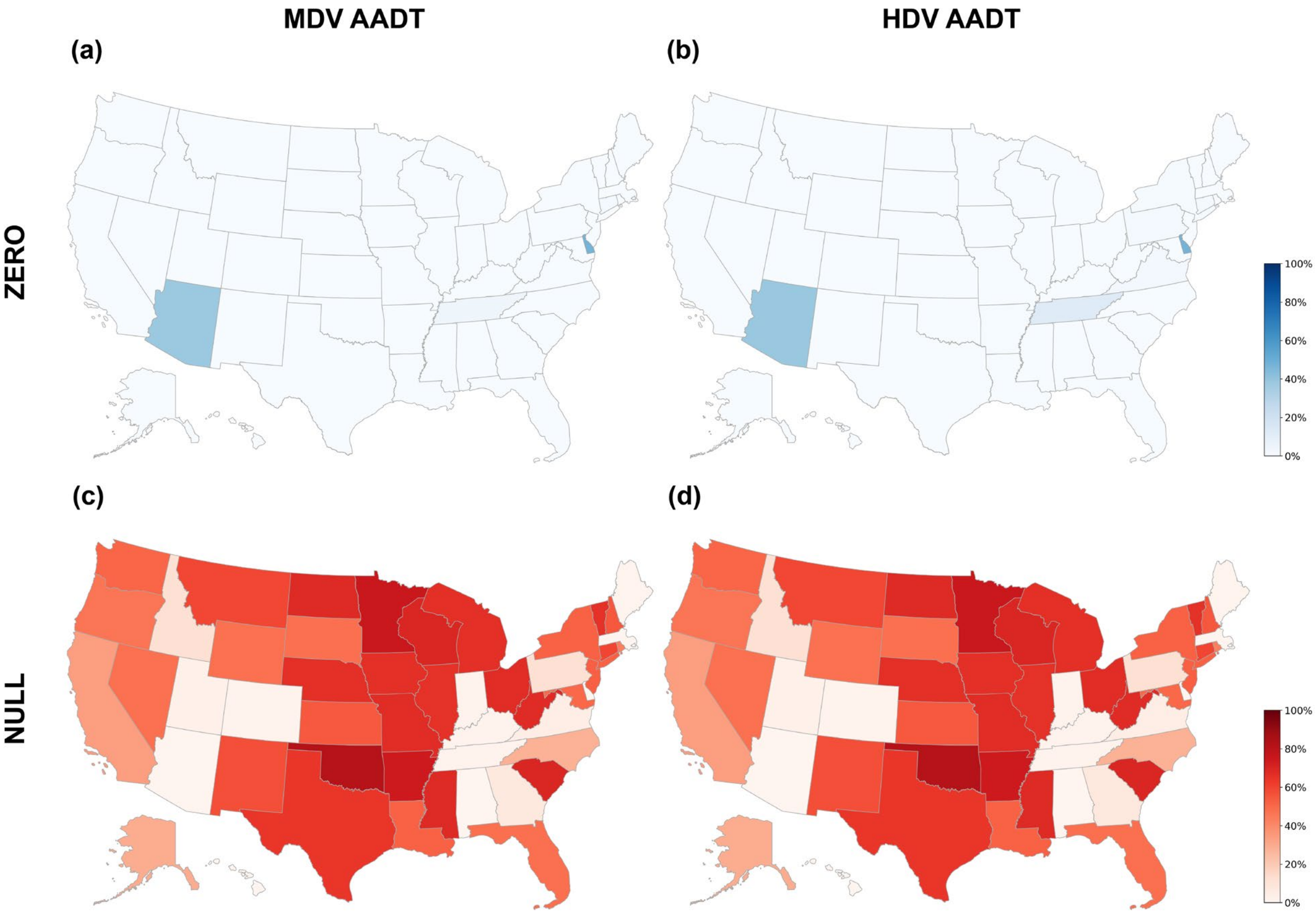

**Fig. 1.** Percentage of HPMS lane-kilometers with zero and null annual average daily traffic (AADT) by vehicle type for each U.S. state. (a) Zero AADT for medium-duty vehicle (MDV) traffic and (b) heavy-duty vehicle (HDV) traffic; (c) Null (or NA) AADT for MDV traffic and (d) HDV traffic.

Two states, Delaware and Arizona, exhibited relatively high proportions of road links with zero MDV or HDV AADT values. Specifically, approximately 47% of lane-kilometers in Delaware and 38% in Arizona had zero MDV and HDV AADT values, compared to an overall state-level mean of about 2%. These states also reported notably low levels of unavailable or 'NA' values for MDV and HDV AADT, with nearly 0% of lane-kilometers lacking data (in contrast to an overall state-level mean of 51% and median of 50%). For this reason, it is assumed that Delaware and Arizona reported zero values in places where AADT data were unavailable, whereas other states classified such instances as 'NA'. To address this, all zero MDV and HDV AADT values in Delaware and Arizona were reclassified as 'NA', affecting 0.6% of lane-kilometers in the HPMS dataset.

The HPMS network also contained duplicate geometries, resulting in overlapping road links. This issue was observed in 26 states on a small scale. At the national level, 1.10% of road links were duplicates (or non-unique geometries), accounting for 3.06% of lane-kilometers in the network. The majority of these duplicate links were identical in all attributes. However, a small number of lane-kilometers (63 lane-kilometers, or 0.0029% of the total lane-kilometers in the network) had duplicate geometries but differed in total AADT, MDV AADT, HDV AADT, or county. To address this issue, we flagged all duplicate geometries and retained only the first occurrence of each in the HPMS dataset, removing the duplicates.

This process resulted in the removal of 0.81% of road links, accounting for 33,447 lane-kilometers (0.83% of lane-kilometers in the network).

### Model Selection and Rationale

Our prior work relied on Linear Regression (LR) to estimate AADT for MDVs and HDVs[18]. However, LR models struggled to accurately estimate traffic volumes for combinations of variables not present in the training set, underscoring the limitations of this approach. Moreover, LR assumes linear relationships among predictors, which may fail to capture the nonlinear complexities inherent in transportation data.

To address the limitations of LR, we use Random Forest Regression (RFR)[37]. RFR is a non-parametric ensemble method that excels at capturing complex, nonlinear relationships between predictor variables and outcomes[38]. Unlike LR, RFR does not rely on predefined assumptions about the data distribution, making it more adaptable to diverse patterns, including sparse and heterogeneous data. Its ability to model interactions without overfitting, combined with its robustness in sparse data scenarios, makes RFR a suitable choice for this task.

RFR works by creating a collection of decision trees, each trained on a random subset of the training data (bootstrapped sample) and a random subset of features at each split[37]. This randomness decorrelates the trees, reducing overfitting and increasing predictive accuracy. Final predictions are made by averaging the outputs of all individual trees in the forest. While this approach may result in some loss of model interpretability, it significantly enhances prediction accuracy. Additionally, RFR is less sensitive to noisy data and outliers and can handle missing values without imputation. RFR is also parallelizable, efficient for large datasets, and requires fewer hyperparameter adjustments compared to other models[39], such as deep learning networks.

Given these advantages, RFR is particularly well-suited for modeling transportation data, which often involves complex, nonlinear relationships and missing values. Its ability to manage noisy, sparse data and capture intricate interactions between predictors makes it an ideal choice for estimating traffic volumes, especially for MDVs and HDVs.

### Machine Learning Model Development and Traffic Volume Estimation

To estimate MDV and HDV AADT on roadway links lacking these data, we developed two national-level RFR models. These models predict AADT for MDV and HDV traffic using the following predictor variables: total AADT, FHWA roadway functional classification, the number of through lanes, and indicator variables for state (FIPS code) and county (FIPS code).

To optimize model performance, we employed Bayesian optimization for hyperparameter tuning using the BayesSearchCV function from the scikit-learn Python package. We used this method to optimize the hyperparameters in order to minimize the root mean squared error (RMSE) over 48 iterations with 3 cross-validation folds. The optimal hyperparameters for each model were identified and are detailed in Table 2.

**Table 2.** Best Hyperparameters for Random Forest Regression Models Optimized via Bayesian Search

| Model | Number of Estimators | Maximum Depth | Minimum Samples Split | Minimum Samples Leaf | Maximum Features |
|---|---|---|---|---|---|
| **MDV AADT** | 120 | None | 2 | 1 | All |
| **HDV AADT** | 115 | 55 | 2 | 1 | All |

*Note. MDV = medium-duty vehicle; HDV = heavy-duty vehicle; AADT = average annual daily traffic.*

Upon examining the partial dependence analysis for the hyperparameters, we found that the optimal values identified through Bayesian optimization did not yield significant improvements in performance compared to the default hyperparameters set by the of the scikit-learn implementation.

Subsequently, we used random forest regression with the default hyperparameters to predict MDV and HDV AADT values for roadway links where these data were unavailable. We then estimated light-duty vehicle (LDV) AADT by subtracting the sum of MDV and HDV AADT from the total AADT for each roadway link.

### Estimating Traffic Density as a Proxy for Traffic-Related Exposure

Building upon the traffic volume estimates derived from the RFR models, we estimated traffic density for each census block in the U.S by vehicle class. Traffic density has previously been used as a surrogate for TRAP exposure[18,19,24] and can potentially be used as a surrogate for other vehicle traffic-related outcomes. Traffic density is calculated following the methodology described in Antonczak et al. (2023)[18] and briefly outlined here.

We obtained 2020 U.S. Census block shapefiles[35] for all 50 states and the District of Columbia. These state-level shapefiles were then merged to create a comprehensive national dataset. To calculate traffic density, we first created a 250-meter spatial buffer around each census block using GIS software. We then intersected the buffered census block with roadway segments that lay entirely within or crossed through the buffer. This buffer accounts for the influence of traffic on roadways near each census block boundary that does not intersect the census block. For example, this captures how air pollution emissions from vehicle traffic occurring near the boundary of a census block would likely affect air quality within the census block. Roadway segments crossing the buffer boundary were divided into portions within and outside the buffer zone. For each road segment, we computed VKT by multiplying AADT by the length of the segment. The VKT for each census block was then calculated by summing the VKT of all roadway segments within the buffered block. The total VKT was then divided by the area of the census block to calculate traffic density, expressed as annual average daily vehicle-kilometers traveled per square kilometer ($VKT/km^2$). This process was repeated for each census block, producing traffic density estimates for each vehicle class and all vehicle classes combined.

### Data Records

Input Data:

The 2018 HPMS dataset, obtained from the BTS, is hosted on our server and can be accessed at https://trafficexposure.uvm.edu/download-hpms.

Census boundaries data can be accessed at https://www2.census.gov/geo/tiger/TIGER/.

Output Data:

Datasets containing the HPMS network with road link average annual daily traffic (AADT) and census block traffic density (VKT/km²) are available on Dryad at ___.

## Technical Validation

We evaluated model performance through several key analyses: aggregate errors using an 80-20 train-test split, residuals analysis, and predictor analysis for both for non-geospatial and spatial factors. Additionally, we performed 5-fold cross-validation and conducted a sensitivity analysis to further assess model robustness. These evaluations help ensure the model effectively captures the relationships between the predictor variables and response variables (MDV AADT and HDV AADT) while minimizing prediction errors and identifying any systematic biases.

### Aggregate Errors

As an initial validity check, we partitioned the HPMS dataset into an 80% training set and a 20% test set. The model evaluation metrics are shown in **Table 3**. Both models performed exceptionally well, with $R^2$ values of 0.99, indicating a strong fit to the data. Additionally, the models exhibited low mean absolute error (MAE) and root mean squared error (RMSE) values. Taken together, these metrics highlight the models' ability to accurately predict traffic volume for both vehicle types across different roadway segments. However, the high $R^2$ values could indicate potential overfitting. To further investigate this, we examined cross-validation scores and sensitivity analysis results.

**Table 3.** Model Performance Metrics for Random Forest Regression Models After Train-Test Split

| Model | $R^2$ | Mean Absolute Error (MAE) | Root Mean Squared Error (RMSE) |
|---|---|---|---|
| MDV AADT | 0.99 | 12.97 | 89.43 |
| HDV AADT | 0.99 | 16.87 | 146.16 |

*Note. MDV = medium-duty vehicle; HDV = heavy-duty vehicle; AADT = average annual daily traffic.*

### Residuals Analysis

In addition to evaluating global accuracy metrics, we conducted a residual analysis by examining residual plots and applying locally weighted scatterplot smoothing (LOWESS) to identify any potential systemic biases in the predictions. The residual plots for the RFR models revealed a nearly flat LOWESS curve, suggesting no significant bias in the predictions across varying levels of traffic volumes (see **Fig. 1**). This finding supports the robustness and generalization ability of the model across different traffic volumes.

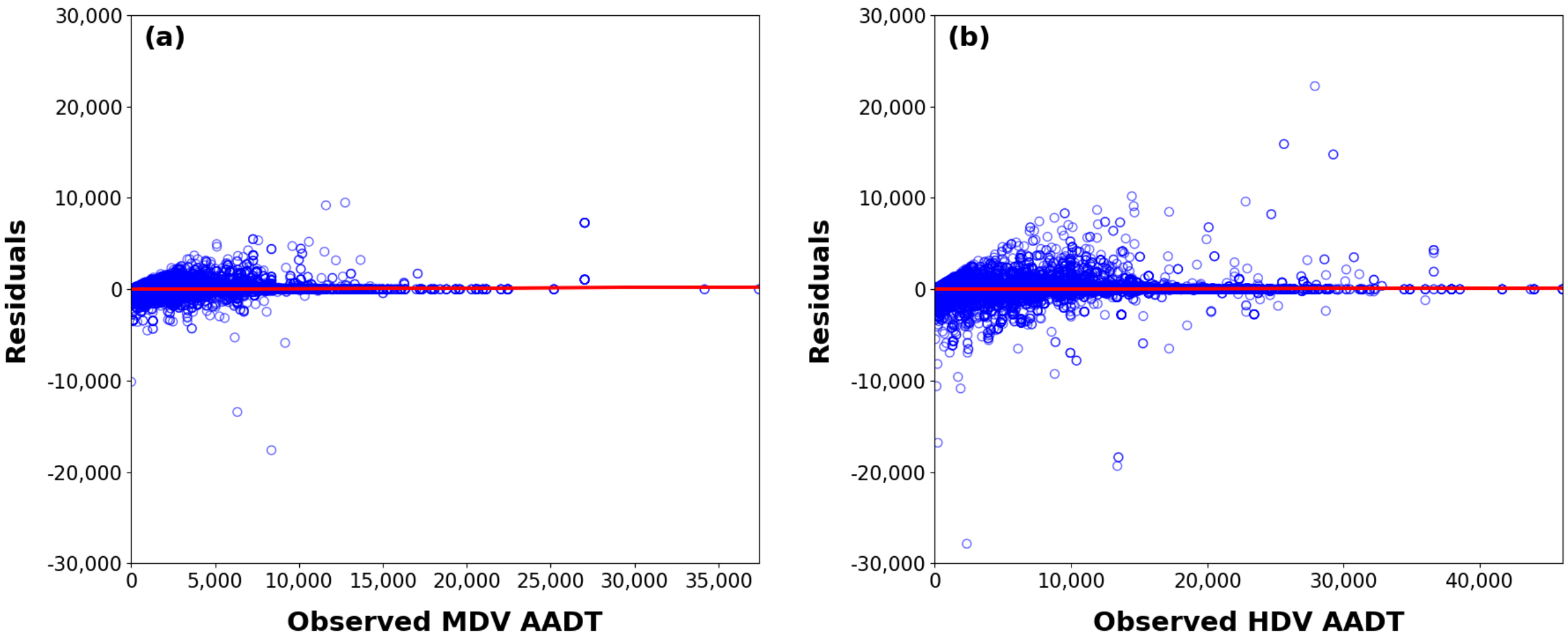

**Fig. 2.** Residuals versus observed annual average daily traffic (AADT). (a) Residuals versus observed medium-duty vehicle (MDV) AADT. (b) Residuals versus observed heavy-duty vehicle (HDV) AADT. The scatter plots show residuals (blue points) plotted against observed AADT values. A LOWESS fit (red line) is applied to the residuals, which closely aligns with the horizontal zero line, indicating minimal bias in the random forest regression model's predictions.

### Non-Geospatial Predictors Analysis

Next, we evaluated the performance of non-geospatial predictors by comparing observed and predicted values for each variable. Box plots illustrating this analysis are shown in **Fig. 2**. The predictions closely align with the observed values, suggesting that there are no systematic under- or over-predictions for any of the non-spatial predictors.

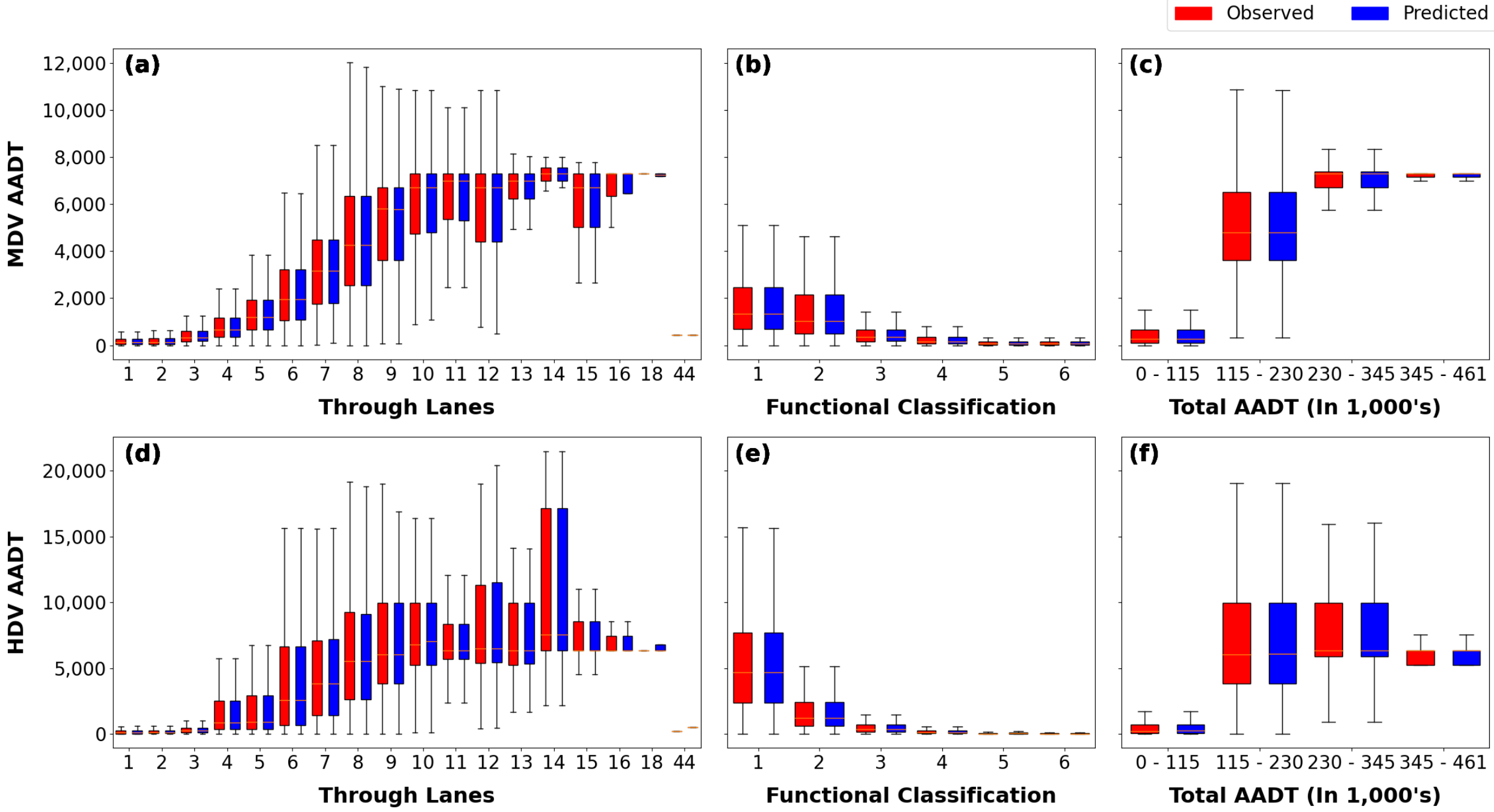

**Fig. 3.** Box plots comparing observed (red) and predicted (blue) annual average daily traffic (AADT) for heavy-duty vehicles (HDVs) and medium-duty vehicles (MDVs) across different random forest model predictor variables. (a) Distribution by the number of through lanes. (b) Distribution by functional classification. (c) Distribution by total AADT (in thousands). The box plots highlight variations in model performance across these predictor variables, emphasizing discrepancies between predicted and observed AADT values. Overall, no significant differences in mean AADT values are observed across the predictor categories.

### Spatial Predictors Analysis

We used the mean absolute percentage error (MAPE) to quantify the accuracy of the model's predictions. MAPE represents the percentage difference between predicted and observed values. To evaluate spatial variability in model performance, we computed the average MAPE for each county in the U.S. based on the 80-20 train-test split. County-level roadway segment MAPEs were estimated as VKT-weighted means to account for variations in roadway segment lengths and mitigate the influence of large absolute errors, which can occur on low-volume roadways (e.g., where traffic volumes are only a few trucks per day).

The resulting county average MAPE values are presented in **Fig. 3.** The average MAPE values for MDV AADT and HDV AADT in Fig 3 are 2.98% and 3.45%, respectively. Most counties exhibit low MAPEs, with many values well below 5%, indicating strong model performance across both urban and rural counties nationwide. However, a very small percentage of counties have MAPE values exceeding 100% (0.06% for MDV AADT and 0.19% for HDV AADT). These counties with high MAPE values are generally rural areas with low-volume roadways, where higher MAPE values are expected due to the smaller traffic volumes.

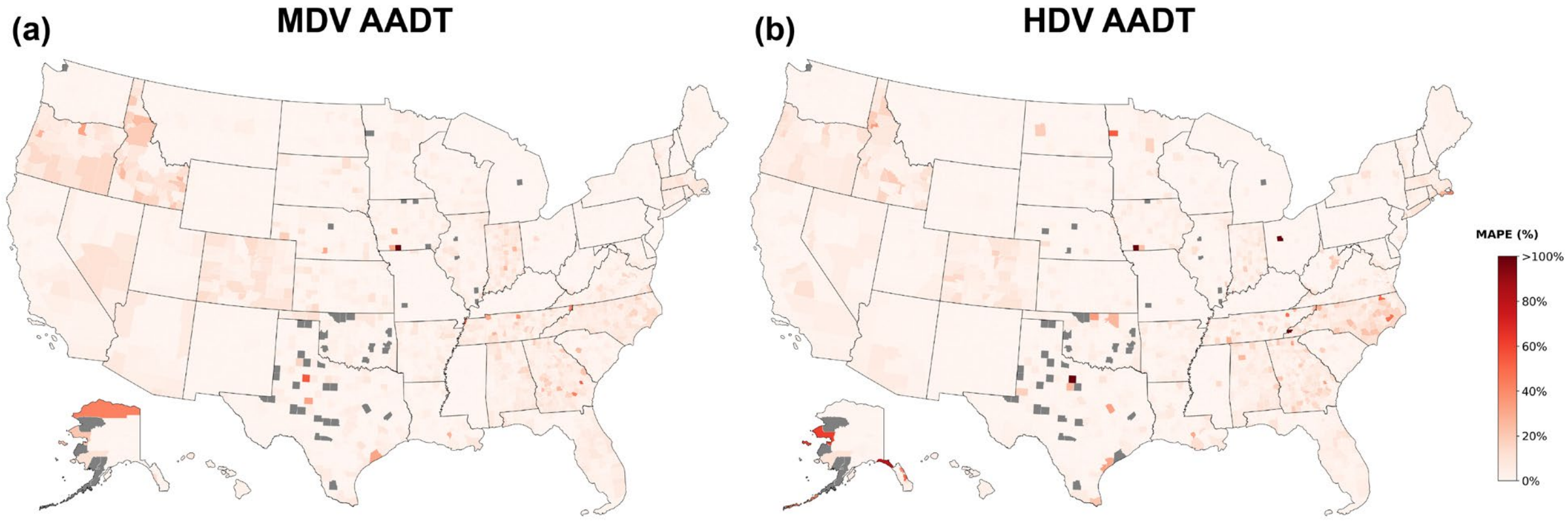

**Fig. 4.** County-level mean absolute percentage error (MAPE) for (a) medium-duty vehicle (MDV) annual average daily traffic (AADT) and (b) heavy-duty vehicle (HDV) AADT, predicted using random forest regression. The MAPE was calculated as the county mean HPMS road link MAPE, weighted by road link vehicle kilometers traveled (VKT) from the 80-20 train-test split. Counties in grey represent those with no observed MDV or HDV AADT data.

**Cross-Validation**

After validating the model's accuracy and assessing potential biases, we performed 5-fold cross-validation to further evaluate model performance. This process provides insight into how the model generalizes across different subsets of the data. The aggregated performance metrics for each model, along with the variance of the outcomes, are shown in **Table 4**. These results demonstrate the robustness of the RFR models, with the variance in the $R^2$ values and MAEs remaining within acceptable thresholds. An $R^2$ of 0.9 or higher is generally considered excellent, indicating a strong correlation between the model's predictions and the actual values.

For both MDV AADT and HDV AADT, the model's MAE and RMSE are small relative to the mean and range of the data. The MAE values account for approximately 2-3% of the mean, indicating that the model’s errors are minimal. Similarly, the RMSE values are small relative to the maximum values of the ranges, constituting only 0.25-0.3% of the maximum value.

**Table 4.** Aggregated Model Performance Metrics for Random Forest Regression Models After 5-Fold Cross-Validation

| **Model** | **$R^2$** | **Mean Absolute Error (MAE)** | **Root Mean Squared Error (RMSE)** |
|---|---|---|---|
| MDV AADT | 0.99 | 14.49 | 98.27 |
| HDV AADT | 0.99 | 18.64 | 149.51 |

*Note. MDV = medium-duty vehicle; HDV = heavy-duty vehicle; AADT = average annual daily traffic.*

**Sensitivity Analysis**

A significant portion of the error in the dataset can be attributed to the unknown error inherent in the HPMS dataset. This includes measurement and bias discrepancies across states, which use different tools and methodologies for traffic data collection. To evaluate the impact of these variations, we introduced known levels Gaussian noise into the data to simulate a range of potential measurement errors and assessed how the RFR models responded to this noise. The sensitivity results, shown in **Fig. 4**, reveal a noticeable, but slight, gradual decrease in model performance when noise was added to the total annual average daily traffic (AADT) predictor. This suggests that while total traffic volume is an important factor, other predictors also influence the final estimates. Additionally, we observed minimal to no impact on the model's performance when noise was introduced into the response variables themselves, further highlighting the robustness of the model's estimations. These findings underscore the model's ability to handle measurement error and its capacity to extract relevant information from a wide range of predictor variables.

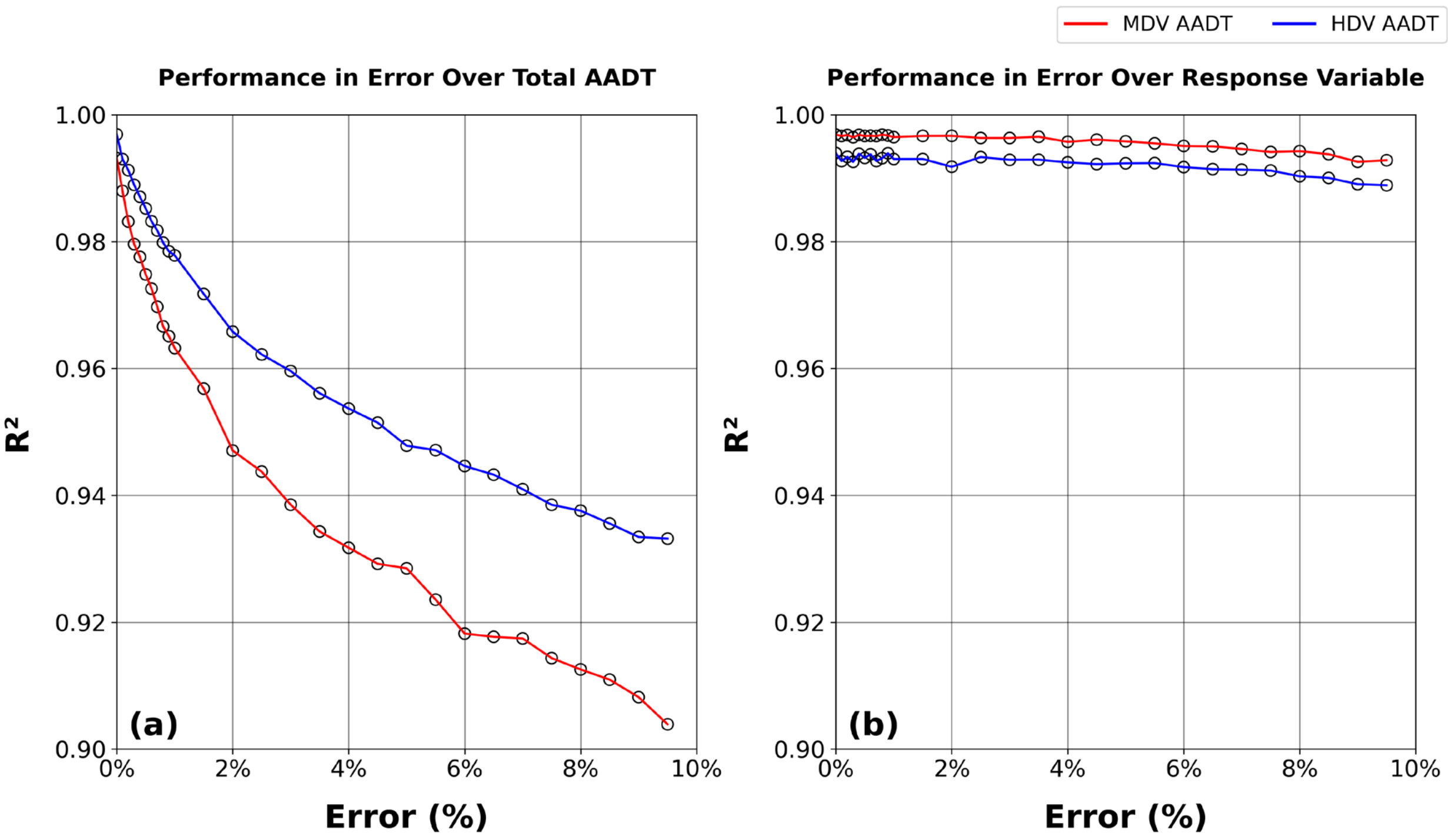

**Fig. 5.** Sensitivity analysis of model performance with increasing error. (a) Performance with respect to total average annual daily traffic (AADT): $R^2$ values plotted against the percentage of error introduced in the total AADT. (b) Performance with respect to response variables: $R^2$ values plotted against the percentage of error introduced in medium-duty vehicle (MDV) and heavy-duty vehicle (HDV) AADT. In both plots, MDV AADT is shown in red, and HDV AADT is shown in blue. Scatter points represent the average $R^2$ for each error level, with model performance assessed across a 0–10% error range.

### Usage Notes

Refer to the README.md document located in the GitHub repository for detailed instructions on how to access, use, and interpret the dataset. For any additional questions or support, please consult the Issues section of the repository or contact the authors via the repository's contact information.

**Assumptions and Limitations:**

The efficacy of the traffic volume estimation model is highly contingent on the quality and accuracy of the underlying HPMS data, as well as the availability of MDV and HDV traffic volume estimates. Variations exist among states regarding the availability of medium- and heavy-duty vehicle traffic volume estimates, influenced by the capability of their counting equipment and specific data requirements. Moreover, the HPMS lacks traffic volume data for low-volume roads in rural areas and minor residential streets in urban areas (i.e., functional classification 7), constituting an estimated 14.8% of the total traffic volume in the U.S.[7] Users should keep in mind that while we show that our models are generally able to accurately estimate MDV and HDV traffic volume where data are missing, there is spatial variation in the level of uncertainty as shown in **Fig 3**. Therefore, users should proceed with caution when using these data for extremely fine-grained geospatial analysis (e.g., evaluating individual roadway segments),

**Code Availability**

The code used for estimating road link-level MDV and HDV AADT on the HPMS network using random forest regression, as well as for estimating LDV AADT based on MDV and HDV AADT, is available upon request at https://github.com/UVM-TRC/HPMS_traffic_data_ML.

The repository also includes scripts for calculating traffic density at the census block level for all vehicle types (LDV, MDV, and HDV), as well as for individual vehicle types. Detailed comments within the code ensure clarity of the modeling process, making it straightforward to adapt the code to other use cases.

**Acknowledgments**

This study was funded by a grant from the National Center for Sustainable Transportation (NCST), supported by the U.S. Department of Transportation (USDOT) through the University Transportation Centers program. This work was also funded by the Environmental Defense Fund, whose work is supported by gifts from Signe Ostby, Scott Cook, and the Valhalla Foundation.

**Author contributions**

The authors confirm contribution to the paper as follows: study conceptualization: GR; data collection: BA; methodology: BA and AC, software (writing, implementations, and debugging): AC, MF, and BA, technical validation: AC and BA; writing – original draft: BA, MF, and AC, writing – review and editing: BA, MF, and GR. All authors reviewed the results and approved the preprint version of the manuscript.

**Competing interests**

The authors declare no competing interests.